\documentclass[pra,twocolumn,showpacs]{revtex4}

\usepackage{graphicx}
\usepackage{hyperref}

\renewcommand{\vec}{\mathbf}

\begin{document}

\title{Dynamics of axialized laser-cooled ions in a Penning trap}
\author{E.S. Phillips}
\author{R.J. Hendricks}
\altaffiliation[Present address: ]{Department for Physics and Astronomy, University of Aarhus, 8000 \AA rhus C., Denmark}
\author{A.M. Abdulla}
\author{H. Ohadi}
\author{D.R. Crick}
\author{K. Koo}
\author{D.M. Segal}
\email[E-mail: ]{D.Segal@imperial.ac.uk}
\author{R.C. Thompson}
\affiliation{Quantum Optics and Laser Science, Blackett Laboratory, Imperial College London, London SW7 2AZ, UK}

\begin{abstract}
We report the experimental characterization of axialization --- a method of reducing the magnetron motion of a small number of ions stored in a Penning trap.  This is an important step in the investigation of the suitability of Penning traps for quantum information processing.  The magnetron motion was coupled to the laser-cooled modified cyclotron motion by the application of a near-resonant oscillating quadrupole potential (the ``axialization drive'').  Measurement of cooling rates of the radial motions of the ions showed an order-of-magnitude increase in the damping rate of the magnetron motion with the axialization drive applied.  The experimental results are in good qualitative agreement with a recent theoretical study.  In particular, a classical avoided crossing was observed in the motional frequencies as the axialization drive frequency was swept through the optimum value, proving that axialization is indeed a resonant effect.
\end{abstract}

\maketitle

\section{Introduction}
The first proposal~\cite{Cirac95a} for the use of trapped ions for quantum information processing (QIP) has been realized in its original and in other forms in several groups around the world~\cite{Schmidtkaler03b, Leibfried03a, Home06a, Haljan05a}.  Recent work on large-scale entanglement and scalability~\cite{Leibfried05a, Haffner05a} of ion-trap quantum computing and the demonstration of the essential feature of quantum error correction~\cite{Chiaverini04a} mean that trapped ions offer one of the most promising systems in which to investigate QIP in the short- to medium-term.  All of the experimental demonstrations to date have been in RF traps of various designs.  Although there are many advantages to using such traps, they have the disadvantages of the motional heating of the ions due to their micromotion, and the relatively strong effects of patch potentials formed on the electrodes, particularly with smaller electrode dimensions.

The use of charged particles in Penning traps for quantum information processing is being investigated by a number of groups~\cite{Mancini00a, Ciaramicoli07a, Galve06a, Koo04a, Castrejon-Pita05a, Porras06a, Bollinger01b, Taylor07a}.  By contrast with RF traps, there is no inherent heating of the ions by the trapping fields since only static fields are employed.  The use of static fields also means that the trap can be designed to have larger trap electrodes, further away from the ions.  Thus, the effect on the trapped particles of patch potentials on the surface of the electrodes should be smaller.  A final advantage of Penning traps is the shielding from external magnetic-field fluctuations through the Meissner effect when a superconducting magnet is employed; magnetic field stability has been found to be a limiting factor in some other experiments~\cite{Riebe06a}.

A major disadvantage of using ions in a Penning trap is that the laser cooling process is greatly complicated by the large Zeeman splitting of the levels (much greater than a typical laser linewidth), and the overall negative energy of one of the two radial motions, the magnetron motion.  This latter problem requires that, in order to ``cool'' the ions, energy must be added to the magnetron motion and simultaneously removed from the other two characteristic motions: the axial motion and the (radial) cyclotron motion.  We have previously demonstrated a technique called axialization for overcoming this problem using laser-cooled magnesium ions~\cite{Powell02a}.  This is a variant on a technique previously used in conjunction with buffer-gas cooling in a Penning trap~\cite{Savard91a}.

Although the hyperfine structure of $^{25}$Mg$^+$ could, in principle, be used for the storage of quantum information, the ground state and the metastable D$_{5/2}$ level of $^{40}$Ca$^+$ have already been shown to be a viable system for this purpose~\cite{Schmidtkaler03c}.  For this reason, and because of the availability of an all-solid-state laser system for addressing the cooling and qubit transitions, we have chosen to investigate the laser cooling of $^{40}$Ca$^+$ ions.  Doppler cooling of this ion in a Penning trap is reported in Ref.~\cite{Koo04a}.  We present here the first demonstration of axialization of this ion in the presence of laser cooling, and the measurement of the damping rates of small numbers of ions, both axialized and unaxialized.

\section{Ion motion in the Penning trap}

The ideal Penning trap has the same geometry as a Paul trap, namely a pair of end-caps and a ring electrode, with all three electrodes having hyperbolic cross-sections.  Traps with non-hyperbolic electrodes can give a reasonable approximation to a pure quadrupole potential and are often employed.  A DC voltage applied between the end-caps and the ring provides confinement in the axial direction but repels the ion in the radial direction.  Radial confinement is provided by adding an axial magnetic field, forcing the ion into cyclotron-like loops.

The motion in the axial direction is simple harmonic motion in a potential well.  The motion in the radial plane is an epicycle, composed of a high-frequency cyclotron motion and a lower frequency drift arising from the cross product of the electric and magnetic fields.  The cyclotron motion is modified from that which would occur in the presence of a magnetic field alone, due to the electric field.  The high- and low-frequency motions are referred to as the modified cyclotron and magnetron motions respectively.

A detailed description of ion motion in a Penning trap may be found in~\cite{Ghosh95a, Major05a}.  In summary, the potential from the hyperbolic electrodes is given by,
\begin{equation}
\phi = U_0(2z^2 - x^2 - y^2) / R^2\;,
\end{equation}
where $U_0$ is the DC voltage and $R^2 = r_0^2 + 2z_0^2$ is a geometrical factor depending on the trap radius ($r_0$) and the end-cap separation ($2z_0$).  The magnetic field, $\vec{B}$, is in the $-z$ direction.  The axial motion is independent of the magnetic field, with frequency,
\begin{equation}
\omega_z = \sqrt{\frac{4eU_0}{mR^2}}\;.
\end{equation}

The radial component of the ion motion may most naturally be examined in a frame rotating with the ions at half the true cyclotron frequency (i.e.\ at $eB/2m$)~\cite{Thompson97a}.  In this frame the force due to the magnetic field is exactly canceled by the Coriolis force.  The two motions are then found to be circular orbits with opposite directions and the same frequency $\omega_1$, where,
\begin{equation}
\omega_1^2 = \omega_c^2 / 4 - \omega_z^2 / 2\;.
\end{equation}
Superpositions of the two motions give rise to elliptical and (for equal amplitudes of the two motions) linear trajectories in the rotating frame.  This is analogous to the way in which circular polarizations of light may be decomposed into linear polarizations and vice versa.

Rotation in the same direction as that of the frame is found to correspond to modified cyclotron motion in the laboratory frame (at $\omega_c/2 + \omega_1$) while rotation in the opposite direction corresponds to magnetron motion (at $\omega_c/2 - 1$).

\section{Laser cooling in a Penning trap}

\begin{figure}
\includegraphics[scale=0.8]{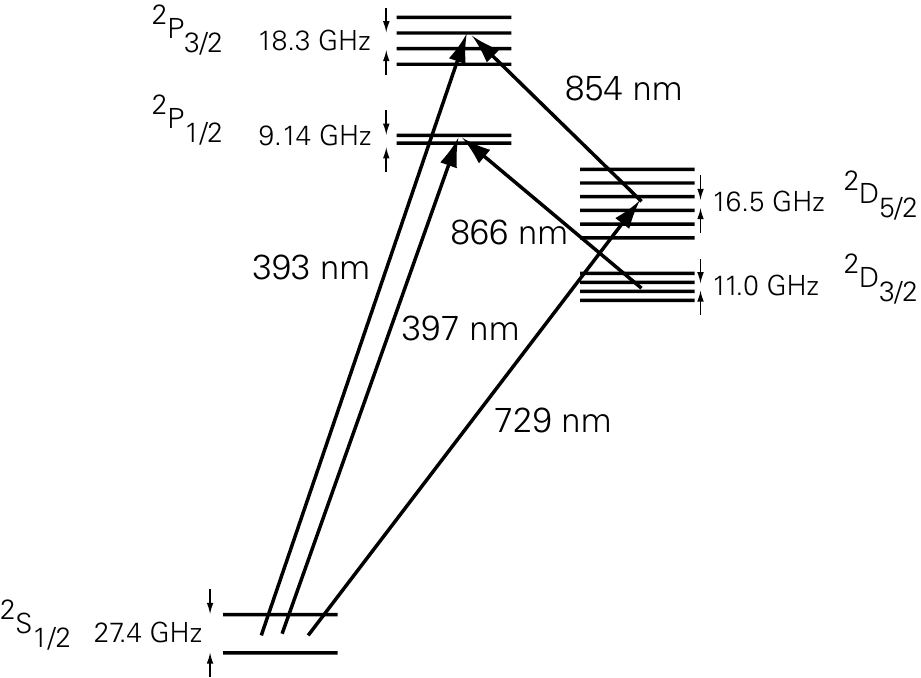}
\caption{\label{fig:calevels}Simplified energy level scheme of $^{40}$Ca$^+$ at a magnetic field of 0.98\,T.}
\end{figure}

Although laser cooling of bound atoms was first demonstrated in a Penning trap~\cite{Wineland78a}, it is rather more difficult than in a Paul trap.  A relatively modest magnetic field of 1\,T can give rise to Zeeman splittings of many gigahertz (see Fig.~\ref{fig:calevels}).  The splittings can be so large that is is impossible to use a single, modulated, laser to address all transitions from a given level.  The set-up and maintenance of many narrow-linewidth, drift-free laser systems is a significant complication.

As well as the large number of lasers required, the choice of detuning from resonance and the (radial) spatial offset of the cooling beam from the center of the trap is also important.  This was investigated in an analytical model in Ref.~\cite{Thompson00a} and further developed in Ref.~\cite{Hendricks07a}.  

\begin{figure}
\includegraphics[scale=0.6]{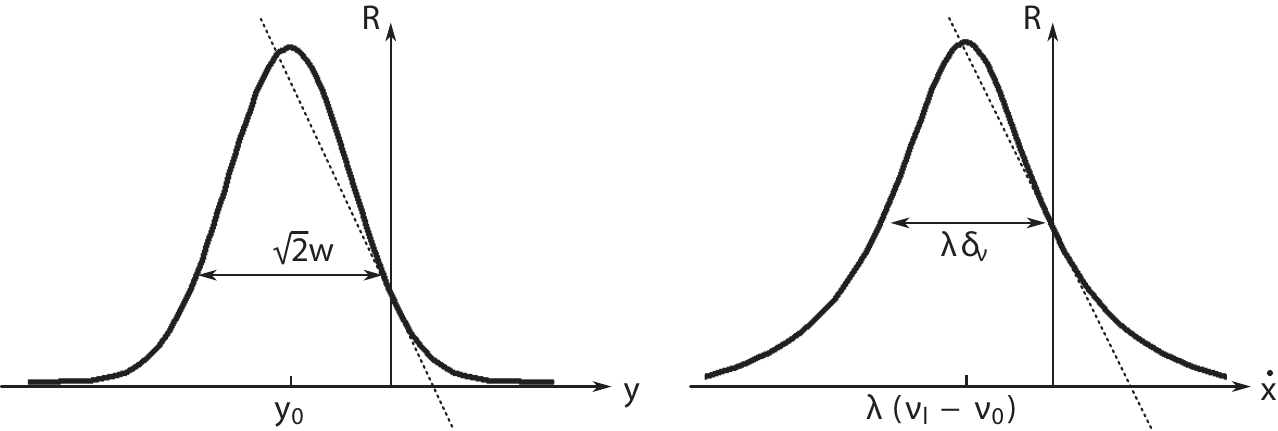}
\caption{\label{fig:lineshapeprofile}The laser cooling efficiency of the radial motions is determined by both the spatial profile of the cooling laser (Gaussian TEM$_{00}$ in the typical case) and the lineshape of the cooling transition (Lorentzian for a cold ion).  The laser scattering rate depends on the instantaneous slopes of the tangents to these two curves and is varied by changing the spatial offset of the beam from trap center and by detuning the laser from resonance.}
\end{figure}

To model the laser cooling process in a Penning trap, it is important to take account of both the lineshape of the cooling transition and also the spatial profile of the cooling laser (Fig.~\ref{fig:lineshapeprofile}).  The ion position in the laser's spatial profile gives rise to a force varying approximately linearly with position along the axis perpendicular to the laser,
\begin{equation}
F_\alpha = -2\alpha m y\;.
\end{equation}
The ion's velocity along the direction of the laser beam leads to a variation of the force on the ion as the ion is Doppler shifted into or out of resonance with the laser frequency, $\nu_L$,
\begin{equation}
F_\beta = -2\beta m \dot{x}\;.
\end{equation}

Adding these two forces to the equations of motion and making the assumptions that $\alpha \ll \omega_z^2$ and $\beta \ll \omega_c$, leads to new solutions with frequencies~\cite{Hendricks07a},
\begin{eqnarray}
\omega_{c'} + i\,\frac{\beta\omega_{c'} - \alpha}{2\omega_1}\;, & & \\
\omega_m + i\,\frac{\alpha - \beta\omega_m}{2\omega_1}\;. & &
\end{eqnarray}

That is, the motional frequencies remain unchanged by the cooling.  The cooling rates (the imaginary parts) are determined by $\alpha$ and $\beta$.  It can be seen that efficient cooling of both motions is not possible due to conflicting requirements: positive $\beta$ (i.e.\ a red-detuned laser) for the modified cyclotron motion and negative $\beta$ (blue-detuned laser) for the magnetron motion.  A compromise may be found where both motions are cooled, but not efficiently~\cite{Itano82a}:
\begin{equation}
\omega_m < \alpha / \beta < \omega_{c'}\;.
\end{equation}
This corresponds to a red-detuned laser and a small spatial offset such that the laser is along the direction of the magnetron motion.

\section{Axialization}

Since it is relatively easy to cool the modified cyclotron motion, it would be desirable to deterministically couple it to the magnetron motion.  In this way the amplitudes of both motions could be reduced without direct damping of the magnetron motion.  This is the essence of the axialization technique.  We have described how the two modes of motion may be thought of as counter-propagating circular motions in a rotating frame.  In order to couple two such modes in a potential well, it is necessary to compress the well along one axis.  This will result in an initially circular motion evolving into an elliptical motion which becomes more and more eccentric until it is linear.  The motion will then become elliptical once again until a circular motion reappears, with equal amplitude but opposite direction to the original motion.

The required compression of the potential well (in the rotating frame) is achieved by adding a static quadrupole field.  To obtain such a quadrupole field in the rotating frame a rotating field must be applied in the laboratory frame.  A quadrupole field oscillating at $\omega_{c}$ can be decomposed into two counter-rotating quadrupole fields each at $\omega_{c}/2$.  In the frame rotating at $\omega_{c}/2$ , one of these rotates at $\omega_{c}$ and has no effect on the motion.  The other is the desired static quadrupole.

An oscillating azimuthal quadrupole field to couple the motions was first proposed by Dehmelt and Wineland~\cite{Wineland75b, Wineland76a}.  Although the motion described is classical, their description of the effect of the coupling field is in quantum mechanical terms.  Such a drive has been applied in the presence of resistive cooling~\cite{Wineland75b} and buffer-gas cooling~\cite{Savard91a} where it is referred to as ``sideband cooling''\footnote{This is unrelated to the laser cooling process of the same name.} or ``radial centering''.

\subsection{Quadrupole drive in the presence of laser cooling}\label{sec:axmodel}

We have recently performed a detailed calculation examining the effect of the quadrupole (axialization) drive when one of the radial motions is laser cooled~\cite{Hendricks07a}.  In the rotating frame, the additional force on the ion due to the static component of the axialization field is given by
\begin{eqnarray}
F_x & = & -\frac{eV_0}{2r_0^2}x\\
F_y & = & +\frac{eV_0}{2r_0^2}y\;,
\end{eqnarray}
where $V_0$ is the amplitude of the axialization drive on the ring segments and $r_0$ is the inner radius of the ring electrode.  We introduce a parameter, $\epsilon$, to describe the axialization strength,
\begin{equation}
\epsilon = \frac{eV_0}{2mr_0^2}\;,
\end{equation}
and write $v = x + iy$.  Thus, we can re-write the force on the ion as
\begin{equation}
F_v = F_x + iF_y = -\epsilon mv^*\;.
\end{equation}

The effect of laser cooling is described in the laboratory frame, with the co-ordinate, $u$:
\begin{equation}
\ddot{u} + (\beta - i\omega_c)\dot{u} - \left(\frac{\omega_z^2}{2} + i\alpha\right) u = 0\;.
\end{equation}
In the rotating frame moving at $\omega_r$, with $v = u \mathrm{e}^{i\omega_r t}$,
\begin{eqnarray}
\nonumber & & \ddot{v} + (\beta - i\omega_c + 2i\omega_r)\dot{v}\\
\nonumber & & \quad - \left(\frac{\omega_z^2}{2} + i\alpha + \omega_r^2 - \omega_{r}(\omega_c + i\beta)\right)v\\
\label{eq:coolax} & & \quad + \epsilon v^* = 0\;.
\end{eqnarray}
where we have included the axialization term.  Rotational symmetry is broken by the $v^*$ term and the solution is in the form of elliptical motion with uniform decay.  This can be written as a superposition of two circular motions in opposite senses and with different amplitudes:
\begin{equation}
v = A\mathrm{e}^{i\omega t} + B\mathrm{e}^{-i\omega^* t}\;.
\end{equation}
The complete solution is described in \cite{Hendricks07a}.  We define $\omega_a$, the axialization-drive frequency and $\Delta$, half the detuning from the true cyclotron frequency, i.e.\ $\omega_a = \omega_c + 2\Delta$.  The solutions to Equation \ref{eq:coolax} have motional frequencies similar to the unperturbed frequencies, $\omega_1$, with a frequency shift $\delta_0$.  The damping rates are obtained from the imaginary part of the solution, $\gamma_0$.
\begin{equation}
\label{eq:perturbedfreq}
\omega = \omega_1 + \delta_0 + i\gamma_0\;.
\end{equation}
The solutions give,
\begin{equation}
\label{eq:delta0}
\delta_0 = \pm \frac{1}{\sqrt{2}}\sqrt{N + \sqrt{N^2 + \Delta^2 M^2}}\;,
\end{equation}
and
\begin{equation}
\gamma_0 = \frac{\beta}{2} + \frac{\Delta M}{2\delta_0}\;.
\end{equation}
The parameters $M$ and $N$ are given by
\begin{equation}
M = (2\alpha - \beta\omega_c) / 2\omega_1\;,
\end{equation}
\begin{equation}
N = \Delta^2 - \frac{M^2}{4} + \frac{|\epsilon|^2}{4\omega_1^2}\;.\label{eq:N}
\end{equation}

$M$ is a measure of the strength of the laser cooling while $N$ is dependent on both the laser cooling strength and the amplitude and detuning of the axialization drive.

Equation \ref{eq:delta0} implies that there are two frequency components in the rotating frame.  Recall that in the laboratory frame, the motional frequencies are $\omega = \omega_r \pm (\omega_1 + \delta_0)$.  There are thus four frequency components observed in the laboratory frame:
\begin{eqnarray}
\omega & = & \omega_{c'} + \Delta + \delta_0\\
\omega & = & \omega_m + \Delta - \delta_0
\end{eqnarray}
with $\delta_0$ taking the two values given by Equation (\ref{eq:delta0}).

We can vary $M$ by changing the cooling laser beam position and detuning from resonance.  $M$ is also affected by the repumper laser detunings as well as the intensities of all the lasers.  These parameters will also affect $N$.  The other free parameters in Equation (\ref{eq:N}) are easily controlled by changing the axialization-drive amplitude and detuning.

\begin{figure}
\includegraphics[scale=1]{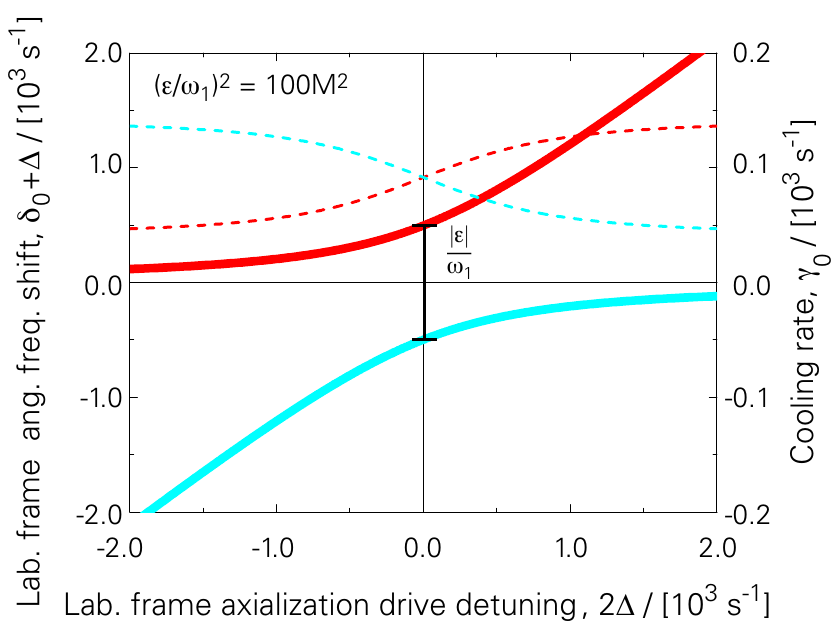}
\caption{\label{fig:eggM2}When $(\epsilon/\omega_1)^2 \gg M^2$, an avoided crossing of the motional frequencies (solid lines) is predicted, shown here near the magnetron frequency.  The damping rates (dashed lines) of the two radial motions are relatively high and insensitive to the axialization-drive frequency.  The red dashed line is the damping rate corresponding to the upper frequency while the blue dashed line corresponds to the lower frequency.}
\end{figure}

We investigated three regimes of axialization-drive amplitude theoretically~\cite{Hendricks07a}.  One of these is particularly important experimentally, where $(\epsilon/\omega_1)^2 \gg M^2$.  In this case the axialization rate is faster than the rate of damping due to laser cooling.  Both the magnetron and modified cyclotron motions are cooled and the sensitivity of the damping rates to the axialization-drive frequency is low compared to the other regimes (Fig.~\ref{fig:eggM2}).  This means that very precise tuning of the axialization drive is unnecessary and that the damping rates are insensitive to magnetic field fluctuations.  An interesting feature of this regime is that an avoided crossing of the motional frequencies measured in the laboratory frame is predicted.

\section{Experimental setup}

The setup has been described in some detail in Refs.~\cite{Koo04a, Hendricks06a}.  We perform laser cooling of small clouds of calcium ions in a Penning trap using diode lasers.  An important feature of the system is that the ring electrode of the trap is divided into quadrants to allow the application of the axialization drive.

\subsection{Laser systems}

A pair of laser diodes in commercial systems (Toptica DL100) provides light on the cooling transition (S$_{1/2}$--P$_{1/2}$, 397\,nm) while home-built extended cavity diode lasers at 866\,nm and 854\,nm (D$_{3/2}$--P$_{1/2}$ and D$_{5/2}$--P$_{3/2}$ respectively) close the cooling cycle.  Repumping from the D$_{5/2}$ level is necessary due to a small amount of population of the P$_{3/2}$ level by the cooling lasers (ASE light at 393\,nm) and subsequent decay to the long-lived D state.

Two lasers are used on the 397\,nm transition on account of the large ground-state Zeeman splitting of 27\,GHz, making a single, modulated source difficult to produce.  The two lasers at 866\,nm each have sidebands imposed on them via an RF source added to the diode current.  The two pairs of sidebands then cover the four $\sigma$ transitions.  A single 854-nm laser is found to be sufficient to repump the small amount of population found in the D$_{5/2}$ level.

All of the diode lasers are locked with side-of-fringe locks to low-finesse cavities.  The cavities are constructed from Zerodur and are tunable via piezo transducers arranged in a re-entrant design to minimize thermal drift.  The cavities are kept in insulated containers which are temperature-stabilized.  Roughly linear drifts of between 15 and 75\,MHz per hour have been measured.  (The cooling transition natural linewidth is 23\,MHz for comparison.)

\subsection{Trap with segmented ring electrode}

\begin{figure}
\includegraphics[scale=1.0]{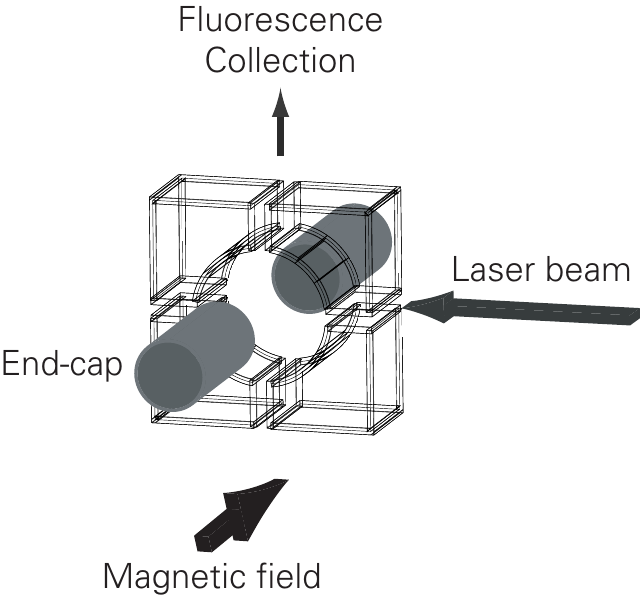}
\caption{\label{fig:trap}Split-ring trap}
\end{figure}

The trap itself (Figure~\ref{fig:trap}) was first described in \cite{vanEijkelenborg99b} and is housed in a vacuum chamber kept at ultra-high vacuum (10$^{-9}$\,mbar or lower).  It approximates an ideal Penning trap but uses non-hyperbolic electrodes with conical cross sections.  The ring electrode is split into four quadrants to allow the application of azimuthal dipolar and quadrupolar potentials.  The magnetic field for ion confinement is provided by an electromagnet with a maximum field of approximately 0.98\,T and a long-term field stability of better than one part in 10$^4$.  With an end-cap voltage, $U_0$, of 4.0\,V, this gives $\omega_z = 2\pi \times 141$\,kHz, $\omega_m = 2\pi \times 23.9$\,kHz, $\omega_{c'} = 2\pi \times 348$\,kHz.  The true cyclotron frequency, $\omega_c$, is $2\pi \times 376$\,kHz.

In the work described below, the axialization drive and the dipolar excitation signal are provided by high-resolution function generators, allowing the determination of relevant frequencies to 1\,Hz or better.  Fluorescence-photon arrival times were measured with a Photomultiplier Tube (PMT) connected to a Time-to-Amplitude Converter (TAC) and digitized by a Multi-Channel Buffer (MCB) in a computer.

\subsection{Measurement of damping rates}

Motional frequencies and damping rates were measured with an established technique~\cite{vanEijkelenborg99a}.  This involves applying an excitation field near-resonant with the motion of interest and performing photon-RF correlation measurements.  As the frequency of the excitation field is swept through the motional resonance, a $\pi$ phase change occurs in the relative phase of the excitation signal and the ionic motion.  This is a feature of any forced damped harmonic oscillator and the width of the phase change yields the damping rate of the motion.  Efficient excitation was achieved by choosing a suitable geometry (dipolar across the ring electrode in the case of radial modes).

\begin{figure}
\includegraphics[scale=1]{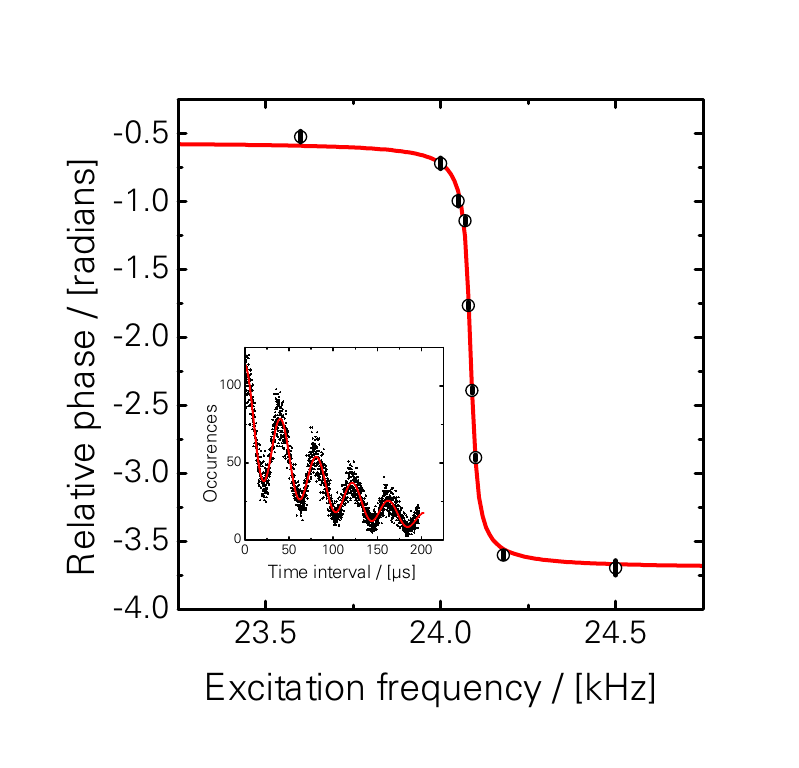}
\caption{\label{fig:photonrf}Typical plot of the phase change as the excitation frequency is stepped across the motional resonance (in this case the magnetron motion).  The fit is an arctan function whose width gives the damping rate.  The motional frequency inferred from the fit shows that there is a shift related to the amplitude of the axialization drive.  \textit{Inset} Sample correlation data from the MCB.  The fit is according to Equation~\ref{eq:photonrf}.  The modulation is due to the ion motion while the initial phase of the modulation depends on the excitation frequency relative to the motional frequency.}
\end{figure}

As the ion moves in the trap, it moves through the gradient of laser beam intensity.  Thus, the probability of emission of a photon is determined by the position of the ion and is related to the phase of the ion's motion.  In this way, the relative phase of the motion with respect to the excitation field may be determined by examining the phase of the photon-RF correlation signal~\cite{vanEijkelenborg99a}.  A typical correlation signal is plotted in the inset of Fig.~\ref{fig:photonrf} and the following equation is fitted,
\begin{equation}
f(\Delta t) = \exp (-a\Delta t) \left[b + |c|\sin(\omega_\mathrm{excit}\Delta t - \phi)\right]\;.
\label{eq:photonrf}
\end{equation}
The underlying exponential decay is due to the normal statistics of delays between photon detection events --- longer delays become progressively less likely.  $\omega_\mathrm{excit}$ is the excitation frequency.  $\phi$ is the most important fit parameter and is the phase difference between the drive and the motion of the ion.

Equation~\ref{eq:photonrf} is fitted to the correlation signals obtained as the excitation frequency is stepped across the motional resonance.  A plot of the phase differences against excitation frequency (Fig.~\ref{fig:photonrf}) then gives two important parameters: the resonant frequency of the motion and the width of the $\pi$ phase change (i.e.\ the damping rate).

\section{Results}

\begin{figure}
\includegraphics[scale=1]{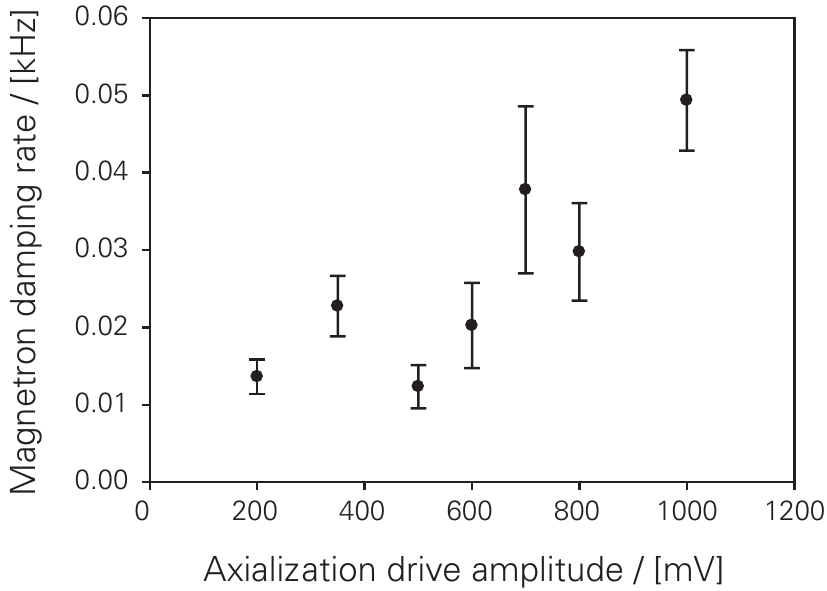}
\caption{\label{fig:axampl}The magnetron damping rate is seen to increase with the axialization drive amplitude for a small drive.  As the two radial motions are more strongly coupled, the otherwise poorly cooled magnetron motion is reduced in amplitude.  For comparison, the damping rate in the unaxialized case was typically less than 5\,Hz.}
\end{figure}

The magnetron damping rate was measured as a function of the axialization drive amplitude.  Measurements were made with small clouds of roughly 10 to 20 ions.  Between measurements, the laser frequencies were re-adjusted to return the fluorescence rate to that recorded at the beginning of the measurement.  This reduced the changes in the damping rate due to laser frequency drifts.  A series of damping rate measurements is shown in Fig.~\ref{fig:axampl}.

The unaxialized magnetron damping rate was typically 2--3\,Hz.  This is lower than expected from theory, in part due to the size of the ion cloud.  With more than one ion, space charge effects increase the size of the cloud and it spends less time in the laser beam.

The magnetron damping rate was seen to increase with increasing axialization-drive amplitude.  The error bars are derived from the fit to the phase plot.  It is difficult to infer the exact relationship between damping rate and axialization drive from the graph since the damping rate is also dependent on the laser frequencies.  Drifts of the four laser-frequency reference cavities were measured to be in the region of tens of megahertz per hour.  A typical damping rate measurement required twenty minutes of data acquisition.

\begin{figure}
\includegraphics[scale=0.7]{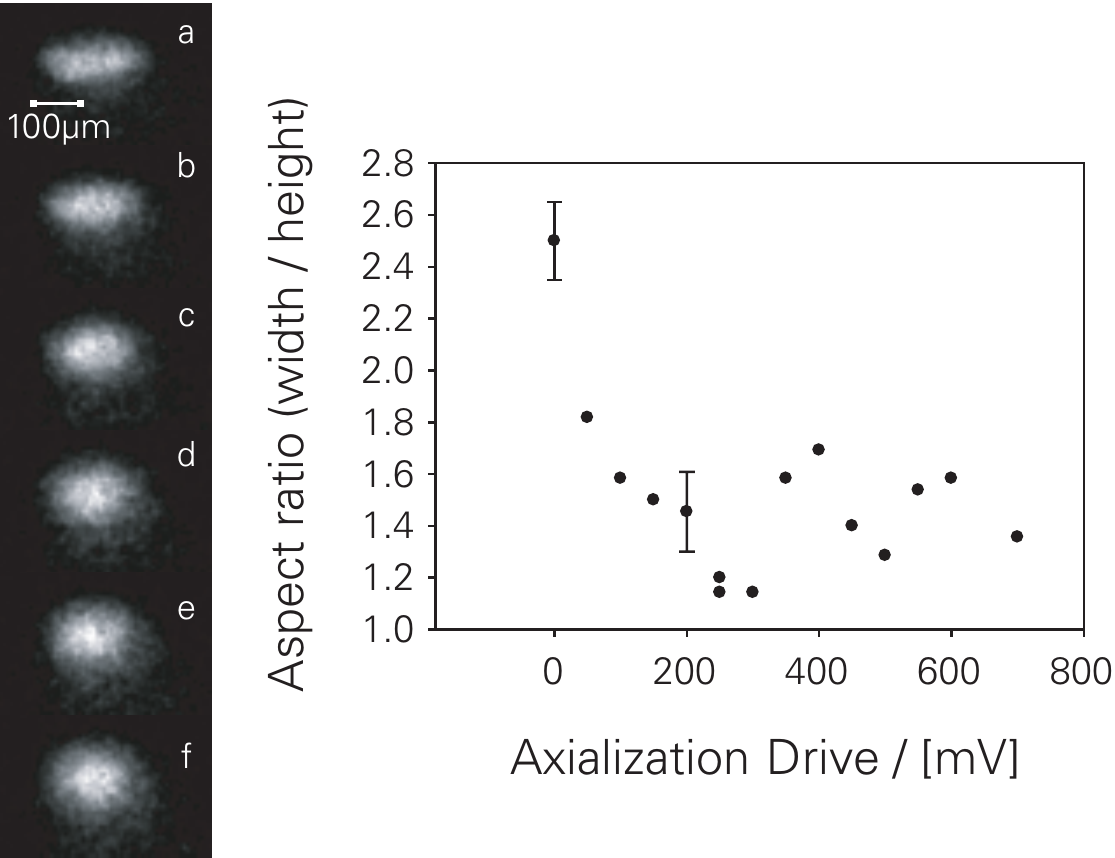}
\caption{\label{fig:aspectratio}The aspect ratio of the cloud rapidly decreases as the axialization drive amplitude is increased, before leveling off above about 200\,mV.  Images \textit{a} through \textit{f} correspond to the first six point in the plot.  The representative error bars are derived from the imaging system resolution.}
\end{figure}

As well as a measured increase in the damping rate, the ion cloud's radius, dominated by the magnetron amplitude, is seen to decrease.  This effect was less marked than in previous experiments with a single magnesium ion~\cite{Powell02a}.  Again, this is due to space-charge effects that come into play when more than one ion is trapped.  The aspect ratio of the ion cloud gives a clearer indication of the effect of the drive for larger numbers of ions (Figure~\ref{fig:aspectratio}).

\subsection{Phase measurements}
In the unaxialized case there is no coupling between the two radial motions, and two $\pi$ phase changes are observed, at the magnetron and modified cyclotron frequencies respectively.  When the coupling is introduced, the frequencies of the resonances are shifted and additional phase changes appear, so that there are observed phase changes at two frequencies near the magnetron frequency and two near the modified cyclotron frequency (Equation~\ref{eq:delta0}).

\begin{figure}
\includegraphics[scale=1]{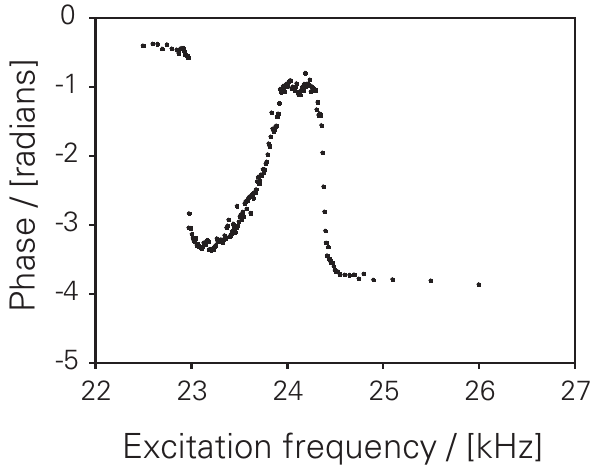}
\caption{\label{fig:avcrossphase}A phase change is observed at two frequencies close to the unmodified frequency of the magnetron motion.  The separation and position are determined by the axialization drive amplitude and detuning from resonance with the true cyclotron frequency.  The shape is in excellent agreement with theoretical predictions~\cite{Hendricks07a}.}
\end{figure}

It is expected that a phase change will occur at each of these ``dressed states''.  An example of the two phase changes near the magnetron motional frequency are shown in Fig.~\ref{fig:avcrossphase}.  A similar pair could also be observed near the modified cyclotron frequency.
		
\subsection{Classical avoided crossing}
The frequencies of the phase changes are predicted by the analytical model summarized in Section~\ref{sec:axmodel}, taking account of the laser cooling rates in a Penning trap and the strength of the axializing field.  We operated in a regime where $(\epsilon/\omega_1)^2\gg M^2$, i.e.\ the coupling rate due to the axializing field was large in comparison to the laser cooling rate.

\begin{figure}
\includegraphics[scale=0.8]{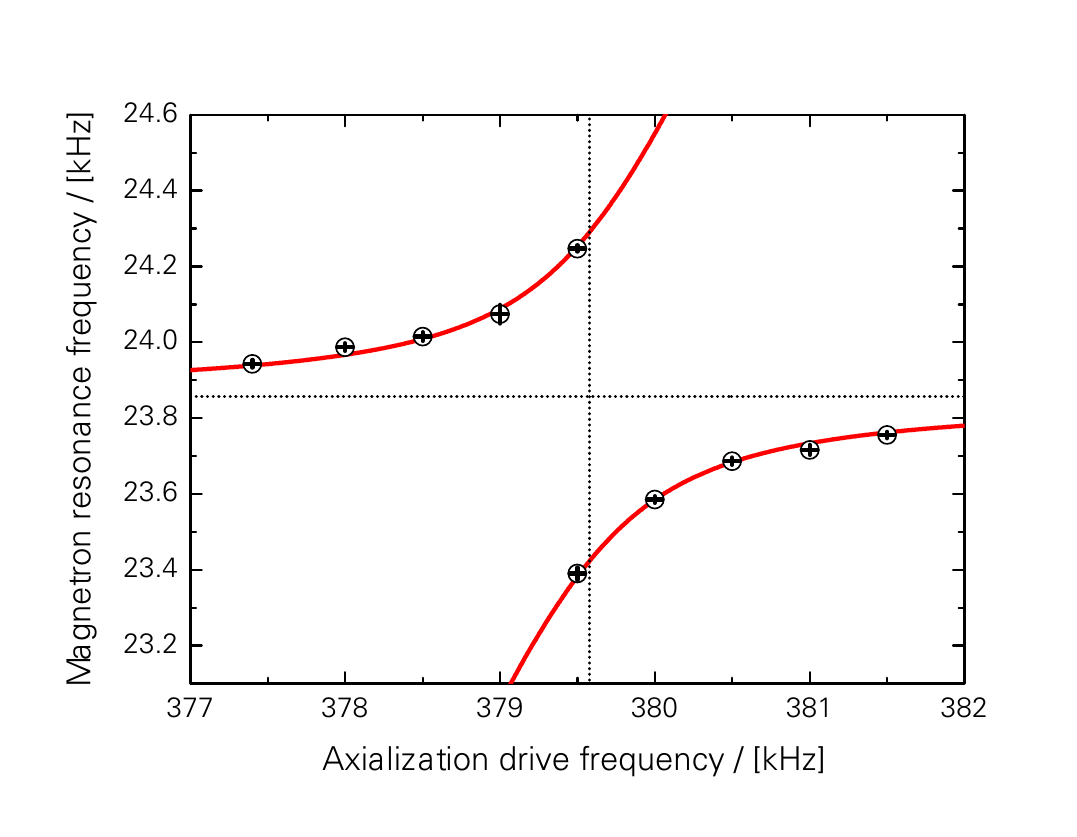}
\caption{\label{fig:avcross}An avoided crossing is observed when the axialization drive frequency is scanned across the true cyclotron frequency.  The fit parameters correspond to a true cyclotron frequency of 379.5\,kHz, an unaxialised magnetron frequency of 23.9\,kHz, and $(\epsilon / \omega_1) = 5.6$\,kHz}
\end{figure}

In this case of a strong axializing field (200\,mV), we expect to see an avoided crossing of the two measured frequencies as we scan the frequency of the axializing field across resonance with the true cyclotron frequency.  We confirmed this by making measurements of the frequencies from plots such as those in Fig.~\ref{fig:avcrossphase} and plotting them against the axialization drive frequency (Figure~\ref{fig:avcross}).  From a fit to the plot, we can infer that the true cyclotron frequency was 379.5\,kHz.  The value for $(\epsilon/\omega_1)$ was $5.6\pm0.2$\,kHz.

From a simulation of the trap geometry in the SIMION~\footnote{SIMION software, Scientific Instrument Services} package, we estimate that the theoretical value for $(\epsilon/\omega_1)$ should be $8.2\pm0.8$\,kHz for a drive of 200\,mV.  The discrepancy between the measured and calculated values is in part due to the measurement being made on a cloud rather than a single ion.

\section{Conclusions}
We have demonstrated the axialization of calcium ions.  The use of RF excitation coupled with time-correlated photon detection allowed the measurement of the damping rates and motional frequencies.  The magnetron damping rate could be increased by an order of magnitude using this technique.  We have also measured the predicted avoided crossing of the magnetron frequency.

The results are in excellent qualitative agreement with theory.  The error in damping rate measurements was relatively large due to laser-frequency drifts.  Improvements to the laser systems should allow quantitative investigation of the relationship between axialization drive amplitude and motional damping rates in the future.

The axialization of laser-cooled calcium ions is an important step in the investigation of the viability of a Penning trap quantum computer.  Future plans include the measurement of heating rates and spectroscopy of the S$_{1/2}$ to D$_{5/2}$ quadrupole transition.  If the expected low heating rates are confirmed, early experiments investigating arrays of Penning traps will allow for scalable quantum information processing with Penning traps~\cite{Castrejon-Pita05a, Galve06a}.

\acknowledgments{The authors gratefully acknowledge the financial support given by the EPSRC under grant GR/R14415/01 and the European Commission through its QUEST Research Training Network, QGATES Research and Technological Development programme and the SCALA Integrated Project.}

\end{document}